\documentstyle [12pt]{article}
\def\fnote#1#2{\begingroup\def\thefootnote{#1}\footnote{#2}\addtocounter
{footnote}{-1}\endgroup}
\evensidemargin=-.25in \textwidth=6.7in \def\half{{\textstyle{1\over2}}}
\def\pl#1{{\sl Phys.~Lett.~\bf #1B}}\def\pr#1{{\sl Phys.~Rev.~\bf D#1}}
\def\cpc#1{{\sl Comp.~Phys. Comm.~\bf #1}}\def\prz#1{{\sl Phys.~Rev.~\bf#1}}
\def\anp#1{{\sl Ann.~Phys.~(NY) \bf #1}}\def\ibid{{\it ibid.}}
\def\etal{{\it et al.}}\def\sqstev{$\sqrt s=40$~TeV}
\def\etatwet{$|\eta|<2.8$}\def\qquds{$q,q'=u,d,s$}\def\ssle{\scriptscriptstyle}
\def\bbar#1{{\llap{\phantom#1}^{\ssle(}\bar#1^{\ssle)}}}

\def\D#1{{\cal D}_#1}\def\braket#1{\langle #1\rangle}
\def\lsim{{<\atop\sim}}\def\Higgs{$H\to\gamma\gamma$}
\def\pptostuff{pp\to q{q^(}'{}^)+n(\gamma)+X}\def\cmssys{in the $pp$ cms
system}
\def\trigcuts{the trigger cuts in (4)}\def\GeV{{\rm GeV}}
\begin{document}
\hfill UTHEP--92--0501 \vskip.01truein \hfill{May 1992}\vskip1truein
\centerline{\Large Multiple photon effects in $pp$ scattering at SSC energies
\fnote{\ast}{Supported in part by the Texas National Research Laboratory
Commission for the Superconducting Super Collider Laboratory via grant
RCFY9201,by the US Department of Energy Contracts DE-FG05-91ER40627
and DE-AC03-76ER00515, and by the Polish Ministry of Education grants
KBN 223729102 and KBN 203809101.}}
\vskip.25truein
\begin{center}
{\sc D.~B. DeLaney\\}{\it Department of Physics and Astronomy\\
The University of Tennessee, Knoxville, TN 37996--1200, USA\\}
{\sc S.~Jadach \fnote{\dagger}{Permanent address:
Institute of Nuclear Physics, ul. Kawiory 26a, Krakow, Poland.}\\}
{\it The University of Tennessee, Knoxville, TN 37996--1200, USA\\ and\\
TH Division, CERN, Geneva 23, Switzerland\\}
{\sc Ch.~Shio and G.~Siopsis\\}{\it Department of Physics and Astronomy\\
The University of Tennessee, Knoxville, TN 37996--1200\\}{\sc B.~F.~L.~Ward \\}
{\it Department of Physics and Astronomy\\
The University of Tennessee, Knoxville, TN 37996--1200, USA\\ and\\
TH Division, CERN, Geneva 23, Switzerland\\ and\\
SLAC, Stanford University, Stanford, CA 94309, USA\\}
\end{center}
\vskip.05truein\baselineskip=21pt
\vskip.25truein\centerline{\bf ABSTRACT}\vskip.1truein\par
The Monte Carlo program SSCYFS2 is used in conjunction with available
parton distribution functions to calculate the effects of multiple photon
radiation on pp scattering at SSC energies. Effects relevant to precision SSC
physics such as Higgs discovery and exploration are illustrated.\par
\renewcommand\thepage{}\vfill\eject
\parskip.1truein \parindent=20pt \pagenumbering{arabic}\par
Now that the SSC is under construction, it is important to prepare for the
maximal exploration of its new energy frontier. Higher-order radiative
corrections to its basic physics processes are then of large significance, for
these corrections determine the precise level at which signals for new physics
or Higgs physics can be separated from background as well as the precise level
at which such signals can be measured. Accordingly, we have recently
initiated~\cite{ddl} the development and implementation of the YFS Monte Carlo
approach in Ref.~\cite{sbm} to higher-order radiative corrections to the SSC
physics processes. In this Letter, we present our results for the multiple
photon radiative effects in $\pptostuff$ where \qquds, and we require that the
$pp$ c.m.s.~production angle of $q(q')$, $\theta_{q(q')}$, must satisfy the SDC
acceptance cut (GEM would have a
similar cut) \etatwet\ for definiteness. (We recall for completeness that the
development in Ref.~\cite{sbm} is based on the original work of Yennie,
Frautschi and Suura in Ref.~\cite{yfs}.)\par
Specifically, we use the Monte Carlo
(MC) event generator SSCYFS2 which was developed in Ref.~\cite{ddl} for
\begin{equation}q\bbar q'\to q\bbar q'+n(\gamma)\label{frst}\end{equation} and
the parton distributions of Refs.~\cite{prt} to simulate, on an event-by-event
basis, the multiple photon initial state radiative effects in
\begin{equation}\pptostuff\ ,\label{secnd}\end{equation} where \qquds.\par
The basic master formula for the cross-section is then the usual parton
distribution convolution
\begin{equation}\sigma=\sum_{q,q'}\int\D{q}(x_1)\D{{q'}}(x_2)\sigma_{
\rm YFS}(x_1x_2s)\,dx_1dx_2\label{thrd}\end{equation} where $\D{q}(x_1)$ is the
usual parton distribution of quark $q$ in $p$ and $\sigma_{\rm YFS}$ is the YFS
multiple-photon cross-section for (1) realized by MC methods in SSCYFS2
in Ref.~\cite{ddl}. We emphasize that the formula in (3) is new
in that it combines the YFS amlpitude based higher order QED corrections
to the reduced hard subprocess with the QCD evolved parton distributions.
The theoretical basis for this is the well-known factorization theorem
for hard hadron--hadron collisions~\cite{mueller}. Equivalently,
since the distributions are strictly defined in the leading-log
approximation framework, each emission of a gluon or a photon from
an incoming parton is independent in that framework so that all gluon emissions
may be factorized away from the photon emissions,as we imply in (3),
for the hard subprocess case.Note that this implies that the QED
corrections to the low energy data from which the $\D{q}$ are evolved
have been done properly~\cite{motsai}.
We emphasize that the entire cross-section in
(3) is also now realized by MC methods by using such methods to choose
$x_1$ and $x_2$ as well as to realize $\sigma_{\rm YFS}$. The resulting program
is called SSCYFSP and it will be described in detail elsewhere~\cite{tap}.
Here, we present results obtained with SSCYFSP and we comment on their
implications for SSC physics objectives.\par
More precisely, our complete trigger cuts for our sample MC data are as
follows:
\begin{equation}E_\gamma>3\ \GeV,\ E_{\rm particle out}>80\ \GeV,\ \theta>\pi/6
\label{frth}\end{equation} so that we expect these data to be relevant to the
GEM and the SDC acceptances. For this trigger, we show in turn in Figs.~1-5 the
photon multiplicity, the total photon transverse momentum, the total photon
mass, the final $v$-distribution of the outgoing $q{q^(}'{}^)$ system, and the
outgoing parton energy fraction distribution \cmssys.\par
What we see in Fig.1 is that the mean value of $n_\gamma$ is .133 $\pm$.369.
Thus, multiple photon effects must be considered in detail in view of our cuts,
where we require $E_\gamma>3$~GeV.\par
In Fig.~2, we show that the total photon transverse momentum has a mean value
\begin{equation}<p_{\bot,tot}>=4.1\pm16.9\ \GeV.\label{ffth}\end{equation} The
key issue regarding background to \Higgs\ in the intermediate regime is how
often we get 40~GeV$\lsim E_\gamma\lsim$75 GeV in the transverse directions. We
see from Fig.~2 that, even allowing for realistic parton distributions, which
clearly degrade substantially the energy available in the reduced collisions in
(3) on the average, we still will have to deal with this question in
detail. Such discussion will appear elsewhere.\par
In Fig.~3, we illustrate further the need to make a detailed study of the
background from multiple photon effects to \Higgs\ via the total photon mass
plot, where we find the mean value\begin{equation}\braket{{M_{n\gamma}}}=
\braket{((\sum_ik_i)^2)^\half }=37.2\pm6.9\ \GeV.\label{sxth}\end{equation}
Again, the value of ${M_{n\gamma}}$ emphasizes that the issue of how many such
$n\gamma$ final states in (2) can fake \Higgs\ has to be studied in
detail.\cite{tap}\par
The $v\equiv(\hat s-\hat s')/\hat s$ distribution shown in Fig.~4 illustrates
again that a substantial fraction of the available energy is radiated away. The
mean value of v is \begin{equation}\braket v=.0179\pm.0668.\label{svnth}
\end{equation} Hence, even for heavy Higgs hunting at the SSC, a detailed
assessment of the effect of this radiation will be required. Such an assessment
will appear elsewhere.\cite{tap}\par
Our final Fig.~5 shows the effect of the interplay of the parton distributions
and our SSCYFSP multiple photon radiation in the final parton energy
distribution. Our YFS radiation shifts the average value of the parton energy
to
lower values by a fraction $\sim0.5\braket v$ of $\sqrt s'/2$ so that the
final parton energy distribution
is only slightly modified to softer values by our YFS radiation.As expected ,
since our reduced cross section scales like $1/s^\prime$,our final parton
energy distribution is indeed significantly softer than our input
distribution: in our input distribution, we have
\begin{equation}\braket{E_q/(\sqrt{s}/2)}\simeq.11\label{eigth}\end{equation}
whereas in Fig.~5 we have
\begin{equation}
\braket{E'_q/(\sqrt{s^\prime}/2)}=.34\pm.27.\label{nnth}\end{equation}
This is consistent with (7) and it illustrates the effect of the reduced
hard cross section on the YFS radiation-the preference for smaller values
of $s^\prime$ weakens the YFS radiation effects relative to what we found
in Ref.~\cite{ddl}, as
expected. This is a consistency check on our work. In all of our
figures, our input distributions were those of Gl\"uck, Reya, and Vogt in
Ref.~\cite{prt}. We have checked that the use of the distributions of Duke and
Owens in Ref.\cite{prt} does not make a significant change in our results. The
trigger cross-section which we find, 5.672$\pm$.002 nanobarns, is consistent
with the results in Ref.~\cite{ddl} and references therein.\par
In summary, we have combined our SSCYFS2 MC event generator with the
parton distributions in Ref.~\cite{prt} to make for the first time a realistic
simulation of $\pptostuff$ at SSC energies on an event-by-event basis. We have
found that the general character of the SSCYFS2 results in Ref.~\cite{ddl} for
$q\bbar q\to q\bbar q'+n(\gamma)$ still hold true for the $pp$ case. Indeed,
our
results emphasize the need for detailed $n(\gamma)$ background studies to
\Higgs\ and $n(\gamma)$ radiative studies in Higgs hunting analysis methods in
general. Such work will appear elsewhere.\cite{tap}\par
$\;$\newline{\bf Acknowledgements }\par
Two of the authors (S.J. and B.F.L.W.) are grateful to Prof.~J.~Ellis for the
kind hospitality of the CERN~TH Division, where part of this work was done.\par
\newpage\flushleft{\Large{\bf Figure Captions}}\par
\noindent {1.} Photon multiplicity for $E_\gamma>3$ GeV in $\pptostuff$,
with \trigcuts, for \sqstev.(In each of our figures, we show histograms
of the respective observable in units as indicated in the title thereof.)\par
\noindent {2.} Total photon transverse momentum in GeV units \cmssys\ for
\trigcuts\ for $\pptostuff$ at \sqstev.\par
\noindent {3.} Total photon squared mass in GeV$^2$ in $\pptostuff$ for
\trigcuts\ at \sqstev.\par
\noindent {4.} $v$-distribution for $\pptostuff$ at \sqstev\ for \trigcuts.\par
\noindent {5.} Final parton energy fraction distribution for $\pptostuff$
at \sqstev\ for \trigcuts; here, the parton energy is measured
in the subprocess cms system.\par
\newpage
\begin{center}
 PHOTON MULTIPLICITY
\end{center}
\setlength{\unitlength}{0.1mm}
\begin{picture}(1600,1500)
\put(0,0){\framebox(1600,1500){ }}
\put(300,250){\begin{picture}(1200,1200)
\put(0,0){\framebox(1200,1200){ }}
\multiput(300.00,0)(300.00,0){4}{\line(0,1){25}}
\multiput(.00,0)(30.00,0){41}{\line(0,1){10}}
\multiput(300.00,1200)(300.00,0){4}{\line(0,-1){25}}
\multiput(.00,1200)(30.00,0){41}{\line(0,-1){10}}
\put(300,-25){\makebox(0,0)[t]{\small $5.000$}}
\put(600,-25){\makebox(0,0)[t]{\small $10.000$}}
\put(900,-25){\makebox(0,0)[t]{\small $15.000$}}
\put(1200,-25){\makebox(0,0)[t]{\small $20.000$}}
\multiput(0,.00)(0,300.00){5}{\line(1,0){25}}
\multiput(0,30.00)(0,30.00){40}{\line(1,0){10}}
\multiput(1200,.00)(0,300.00){5}{\line(-1,0){25}}
\multiput(1200,30.00)(0,30.00){40}{\line(-1,0){10}}
\put(-25,0){\makebox(0,0)[r]{\small $.000\cdot10^{5}$}}
\put(-25,300){\makebox(0,0)[r]{\small $2.500\cdot10^{5}$}}
\put(-25,600){\makebox(0,0)[r]{\small $5.000\cdot10^{5}$}}
\put(-25,900){\makebox(0,0)[r]{\small $7.500\cdot10^{5}$}}
\put(-25,1200){\makebox(0,0)[r]{\small $10.000\cdot10^{5}$}}
\end{picture}}
\put(300,250){\begin{picture}(1200,1200)
\thinlines
\newcommand{\x}[3]{\put(#1,#2){\line(1,0){#3}}}
\newcommand{\y}[3]{\put(#1,#2){\line(0,1){#3}}}
\newcommand{\z}[3]{\put(#1,#2){\line(0,-1){#3}}}
\newcommand{\e}[3]{\put(#1,#2){\line(0,1){#3}}}
\y{0}{0}{1052}\x{0}{1052}{59}
\z{59}{1052}{916}\x{59}{136}{60}
\z{119}{136}{126}\x{119}{10}{60}
\z{179}{10}{10}\x{179}{0}{60}
\y{239}{0}{0}\x{239}{0}{60}
\y{299}{0}{0}\x{299}{0}{60}
\y{359}{0}{0}\x{359}{0}{60}
\y{419}{0}{0}\x{419}{0}{60}
\y{479}{0}{0}\x{479}{0}{60}
\y{539}{0}{0}\x{539}{0}{60}
\y{599}{0}{0}\x{599}{0}{60}
\y{659}{0}{0}\x{659}{0}{60}
\y{719}{0}{0}\x{719}{0}{60}
\y{779}{0}{0}\x{779}{0}{60}
\y{839}{0}{0}\x{839}{0}{60}
\y{899}{0}{0}\x{899}{0}{60}
\y{959}{0}{0}\x{959}{0}{60}
\y{1019}{0}{0}\x{1019}{0}{60}
\y{1079}{0}{0}\x{1079}{0}{60}
\y{1139}{0}{0}\x{1139}{0}{60}
\end{picture}}
\end{picture}
\newpage
\begin{center}
 PHOTON SUM TRANSVERSE MOMENTUM (GeV)
\end{center}
\setlength{\unitlength}{0.1mm}
\begin{picture}(1600,1500)
\put(0,0){\framebox(1600,1500){ }}
\put(300,250){\begin{picture}(1200,1200)
\put(0,0){\framebox(1200,1200){ }}
\multiput(300.00,0)(300.00,0){4}{\line(0,1){25}}
\multiput(.00,0)(30.00,0){41}{\line(0,1){10}}
\multiput(300.00,1200)(300.00,0){4}{\line(0,-1){25}}
\multiput(.00,1200)(30.00,0){41}{\line(0,-1){10}}
\put(300,-25){\makebox(0,0)[t]{\small $.500\cdot10^{2}$}}
\put(600,-25){\makebox(0,0)[t]{\small $1.000\cdot10^{2}$}}
\put(900,-25){\makebox(0,0)[t]{\small $1.500\cdot10^{2}$}}
\put(1200,-25){\makebox(0,0)[t]{\small $2.000\cdot10^{2}$}}
\multiput(0,.00)(0,300.00){5}{\line(1,0){25}}
\multiput(0,30.00)(0,30.00){40}{\line(1,0){10}}
\multiput(1200,.00)(0,300.00){5}{\line(-1,0){25}}
\multiput(1200,30.00)(0,30.00){40}{\line(-1,0){10}}
\put(-25,0){\makebox(0,0)[r]{\small $.000\cdot10^{4}$}}
\put(-25,300){\makebox(0,0)[r]{\small $2.500\cdot10^{4}$}}
\put(-25,600){\makebox(0,0)[r]{\small $5.000\cdot10^{4}$}}
\put(-25,900){\makebox(0,0)[r]{\small $7.500\cdot10^{4}$}}
\put(-25,1200){\makebox(0,0)[r]{\small $10.000\cdot10^{4}$}}
\end{picture}}
\put(300,250){\begin{picture}(1200,1200)
\thinlines
\newcommand{\x}[3]{\put(#1,#2){\line(1,0){#3}}}
\newcommand{\y}[3]{\put(#1,#2){\line(0,1){#3}}}
\newcommand{\z}[3]{\put(#1,#2){\line(0,-1){#3}}}
\newcommand{\e}[3]{\put(#1,#2){\line(0,1){#3}}}
\y{0}{0}{1200}\x{0}{1200}{39}
\z{39}{1200}{1122}\x{39}{78}{40}
\z{79}{78}{44}\x{79}{34}{40}
\z{119}{34}{15}\x{119}{19}{40}
\z{159}{19}{7}\x{159}{12}{40}
\z{199}{12}{3}\x{199}{9}{40}
\z{239}{9}{3}\x{239}{6}{40}
\z{279}{6}{2}\x{279}{4}{40}
\z{319}{4}{1}\x{319}{3}{40}
\z{359}{3}{1}\x{359}{2}{40}
\y{399}{2}{0}\x{399}{2}{40}
\z{439}{2}{1}\x{439}{1}{40}
\y{479}{1}{0}\x{479}{1}{40}
\y{519}{1}{0}\x{519}{1}{40}
\z{559}{1}{1}\x{559}{0}{40}
\y{599}{0}{1}\x{599}{1}{40}
\z{639}{1}{1}\x{639}{0}{40}
\y{679}{0}{0}\x{679}{0}{40}
\y{719}{0}{0}\x{719}{0}{40}
\y{759}{0}{0}\x{759}{0}{40}
\y{799}{0}{0}\x{799}{0}{40}
\y{839}{0}{0}\x{839}{0}{40}
\y{879}{0}{0}\x{879}{0}{40}
\y{919}{0}{0}\x{919}{0}{40}
\y{959}{0}{0}\x{959}{0}{40}
\y{999}{0}{0}\x{999}{0}{40}
\y{1039}{0}{0}\x{1039}{0}{40}
\y{1079}{0}{0}\x{1079}{0}{40}
\y{1119}{0}{0}\x{1119}{0}{40}
\y{1159}{0}{0}\x{1159}{0}{40}
\end{picture}}
\end{picture}
\newpage
\begin{center}
 PHOTON SUM MASS SQUARED (GeV$^2$)
\end{center}
\setlength{\unitlength}{0.1mm}
\begin{picture}(1600,1500)
\put(0,0){\framebox(1600,1500){ }}
\put(300,250){\begin{picture}(1200,1200)
\put(0,0){\framebox(1200,1200){ }}
\multiput(300.00,0)(300.00,0){4}{\line(0,1){25}}
\multiput(.00,0)(30.00,0){41}{\line(0,1){10}}
\multiput(300.00,1200)(300.00,0){4}{\line(0,-1){25}}
\multiput(.00,1200)(30.00,0){41}{\line(0,-1){10}}
\put(300,-25){\makebox(0,0)[t]{\small $1.000\cdot10^{4}$}}
\put(600,-25){\makebox(0,0)[t]{\small $2.000\cdot10^{4}$}}
\put(900,-25){\makebox(0,0)[t]{\small $3.000\cdot10^{4}$}}
\put(1200,-25){\makebox(0,0)[t]{\small $4.000\cdot10^{4}$}}
\multiput(0,.00)(0,300.00){5}{\line(1,0){25}}
\multiput(0,30.00)(0,30.00){40}{\line(1,0){10}}
\multiput(1200,.00)(0,300.00){5}{\line(-1,0){25}}
\multiput(1200,30.00)(0,30.00){40}{\line(-1,0){10}}
\put(-25,0){\makebox(0,0)[r]{\small $.000\cdot10^{4}$}}
\put(-25,300){\makebox(0,0)[r]{\small $2.500\cdot10^{4}$}}
\put(-25,600){\makebox(0,0)[r]{\small $5.000\cdot10^{4}$}}
\put(-25,900){\makebox(0,0)[r]{\small $7.500\cdot10^{4}$}}
\put(-25,1200){\makebox(0,0)[r]{\small $10.000\cdot10^{4}$}}
\end{picture}}
\put(300,250){\begin{picture}(1200,1200)
\thinlines
\newcommand{\x}[3]{\put(#1,#2){\line(1,0){#3}}}
\newcommand{\y}[3]{\put(#1,#2){\line(0,1){#3}}}
\newcommand{\z}[3]{\put(#1,#2){\line(0,-1){#3}}}
\newcommand{\e}[3]{\put(#1,#2){\line(0,1){#3}}}
\y{0}{0}{1200}\x{0}{1200}{39}
\z{39}{1200}{1191}\x{39}{9}{40}
\z{79}{9}{5}\x{79}{4}{40}
\z{119}{4}{2}\x{119}{2}{40}
\z{159}{2}{1}\x{159}{1}{40}
\y{199}{1}{0}\x{199}{1}{40}
\z{239}{1}{1}\x{239}{0}{40}
\y{279}{0}{0}\x{279}{0}{40}
\y{319}{0}{0}\x{319}{0}{40}
\y{359}{0}{0}\x{359}{0}{40}
\y{399}{0}{0}\x{399}{0}{40}
\y{439}{0}{0}\x{439}{0}{40}
\y{479}{0}{0}\x{479}{0}{40}
\y{519}{0}{0}\x{519}{0}{40}
\y{559}{0}{0}\x{559}{0}{40}
\y{599}{0}{0}\x{599}{0}{40}
\y{639}{0}{0}\x{639}{0}{40}
\y{679}{0}{0}\x{679}{0}{40}
\y{719}{0}{0}\x{719}{0}{40}
\y{759}{0}{0}\x{759}{0}{40}
\y{799}{0}{0}\x{799}{0}{40}
\y{839}{0}{0}\x{839}{0}{40}
\y{879}{0}{0}\x{879}{0}{40}
\y{919}{0}{0}\x{919}{0}{40}
\y{959}{0}{0}\x{959}{0}{40}
\y{999}{0}{0}\x{999}{0}{40}
\y{1039}{0}{0}\x{1039}{0}{40}
\y{1079}{0}{0}\x{1079}{0}{40}
\y{1119}{0}{0}\x{1119}{0}{40}
\y{1159}{0}{0}\x{1159}{0}{40}
\end{picture}}
\end{picture}
\newpage
\begin{center}
$v$--DISTRIBUTION
\end{center}
\setlength{\unitlength}{0.1mm}
\begin{picture}(1600,1500)
\put(0,0){\framebox(1600,1500){ }}
\put(300,250){\begin{picture}(1200,1200)
\put(0,0){\framebox(1200,1200){ }}
\multiput(300.00,0)(300.00,0){4}{\line(0,1){25}}
\multiput(.00,0)(30.00,0){41}{\line(0,1){10}}
\multiput(300.00,1200)(300.00,0){4}{\line(0,-1){25}}
\multiput(.00,1200)(30.00,0){41}{\line(0,-1){10}}
\put(300,-25){\makebox(0,0)[t]{\small $2.500\cdot10^{-1}$}}
\put(600,-25){\makebox(0,0)[t]{\small $5.000\cdot10^{-1}$}}
\put(900,-25){\makebox(0,0)[t]{\small $7.500\cdot10^{-1}$}}
\put(1200,-25){\makebox(0,0)[t]{\small $10.000\cdot10^{-1}$}}
\multiput(0,.00)(0,300.00){5}{\line(1,0){25}}
\multiput(0,30.00)(0,30.00){40}{\line(1,0){10}}
\multiput(1200,.00)(0,300.00){5}{\line(-1,0){25}}
\multiput(1200,30.00)(0,30.00){40}{\line(-1,0){10}}
\put(-25,0){\makebox(0,0)[r]{\small $.000\cdot10^{4}$}}
\put(-25,300){\makebox(0,0)[r]{\small $2.500\cdot10^{4}$}}
\put(-25,600){\makebox(0,0)[r]{\small $5.000\cdot10^{4}$}}
\put(-25,900){\makebox(0,0)[r]{\small $7.500\cdot10^{4}$}}
\put(-25,1200){\makebox(0,0)[r]{\small $10.000\cdot10^{4}$}}
\end{picture}}
\put(300,250){\begin{picture}(1200,1200)
\thinlines
\newcommand{\x}[3]{\put(#1,#2){\line(1,0){#3}}}
\newcommand{\y}[3]{\put(#1,#2){\line(0,1){#3}}}
\newcommand{\z}[3]{\put(#1,#2){\line(0,-1){#3}}}
\newcommand{\e}[3]{\put(#1,#2){\line(0,1){#3}}}
\y{0}{0}{296}\x{0}{296}{39}
\y{39}{296}{3}\x{39}{299}{40}
\z{79}{299}{125}\x{79}{174}{40}
\z{119}{174}{52}\x{119}{122}{40}
\z{159}{122}{28}\x{159}{94}{40}
\z{199}{94}{18}\x{199}{76}{40}
\z{239}{76}{12}\x{239}{64}{40}
\z{279}{64}{8}\x{279}{56}{40}
\z{319}{56}{6}\x{319}{50}{40}
\z{359}{50}{4}\x{359}{46}{40}
\z{399}{46}{4}\x{399}{42}{40}
\z{439}{42}{3}\x{439}{39}{40}
\z{479}{39}{1}\x{479}{38}{40}
\z{519}{38}{2}\x{519}{36}{40}
\y{559}{36}{0}\x{559}{36}{40}
\z{599}{36}{36}\x{599}{0}{40}
\y{639}{0}{0}\x{639}{0}{40}
\y{679}{0}{0}\x{679}{0}{40}
\y{719}{0}{0}\x{719}{0}{40}
\y{759}{0}{0}\x{759}{0}{40}
\y{799}{0}{0}\x{799}{0}{40}
\y{839}{0}{0}\x{839}{0}{40}
\y{879}{0}{0}\x{879}{0}{40}
\y{919}{0}{0}\x{919}{0}{40}
\y{959}{0}{0}\x{959}{0}{40}
\y{999}{0}{0}\x{999}{0}{40}
\y{1039}{0}{0}\x{1039}{0}{40}
\y{1079}{0}{0}\x{1079}{0}{40}
\y{1119}{0}{0}\x{1119}{0}{40}
\y{1159}{0}{0}\x{1159}{0}{40}
\end{picture}}
\end{picture}
\newpage
\begin{center}
 PARTON ENERGY FRACTION
\end{center}
\setlength{\unitlength}{0.1mm}
\begin{picture}(1600,1500)
\put(0,0){\framebox(1600,1500){ }}
\put(300,250){\begin{picture}(1200,1200)
\put(0,0){\framebox(1200,1200){ }}
\multiput(300.00,0)(300.00,0){4}{\line(0,1){25}}
\multiput(.00,0)(30.00,0){41}{\line(0,1){10}}
\multiput(300.00,1200)(300.00,0){4}{\line(0,-1){25}}
\multiput(.00,1200)(30.00,0){41}{\line(0,-1){10}}
\put(300,-25){\makebox(0,0)[t]{\small $2.500\cdot10^{-1}$}}
\put(600,-25){\makebox(0,0)[t]{\small $5.000\cdot10^{-1}$}}
\put(900,-25){\makebox(0,0)[t]{\small $7.500\cdot10^{-1}$}}
\put(1200,-25){\makebox(0,0)[t]{\small $10.000\cdot10^{-1}$}}
\multiput(0,.00)(0,300.00){5}{\line(1,0){25}}
\multiput(0,30.00)(0,30.00){40}{\line(1,0){10}}
\multiput(1200,.00)(0,300.00){5}{\line(-1,0){25}}
\multiput(1200,30.00)(0,30.00){40}{\line(-1,0){10}}
\put(-25,0){\makebox(0,0)[r]{\small $.000\cdot10^{4}$}}
\put(-25,300){\makebox(0,0)[r]{\small $2.500\cdot10^{4}$}}
\put(-25,600){\makebox(0,0)[r]{\small $5.000\cdot10^{4}$}}
\put(-25,900){\makebox(0,0)[r]{\small $7.500\cdot10^{4}$}}
\put(-25,1200){\makebox(0,0)[r]{\small $10.000\cdot10^{4}$}}
\end{picture}}
\put(300,250){\begin{picture}(1200,1200)
\thinlines
\newcommand{\x}[3]{\put(#1,#2){\line(1,0){#3}}}
\newcommand{\y}[3]{\put(#1,#2){\line(0,1){#3}}}
\newcommand{\z}[3]{\put(#1,#2){\line(0,-1){#3}}}
\newcommand{\e}[3]{\put(#1,#2){\line(0,1){#3}}}
\y{0}{0}{319}\x{0}{319}{39}
\y{39}{319}{324}\x{39}{643}{40}
\z{79}{643}{91}\x{79}{552}{40}
\z{119}{552}{95}\x{119}{457}{40}
\z{159}{457}{70}\x{159}{387}{40}
\z{199}{387}{55}\x{199}{332}{40}
\z{239}{332}{39}\x{239}{293}{40}
\z{279}{293}{34}\x{279}{259}{40}
\z{319}{259}{26}\x{319}{233}{40}
\z{359}{233}{21}\x{359}{212}{40}
\z{399}{212}{18}\x{399}{194}{40}
\z{439}{194}{15}\x{439}{179}{40}
\z{479}{179}{14}\x{479}{165}{40}
\z{519}{165}{10}\x{519}{155}{40}
\z{559}{155}{12}\x{559}{143}{40}
\z{599}{143}{6}\x{599}{137}{40}
\z{639}{137}{9}\x{639}{128}{40}
\z{679}{128}{7}\x{679}{121}{40}
\z{719}{121}{7}\x{719}{114}{40}
\z{759}{114}{5}\x{759}{109}{40}
\z{799}{109}{5}\x{799}{104}{40}
\z{839}{104}{6}\x{839}{98}{40}
\z{879}{98}{5}\x{879}{93}{40}
\z{919}{93}{3}\x{919}{90}{40}
\z{959}{90}{3}\x{959}{87}{40}
\z{999}{87}{2}\x{999}{85}{40}
\z{1039}{85}{5}\x{1039}{80}{40}
\z{1079}{80}{2}\x{1079}{78}{40}
\z{1119}{78}{3}\x{1119}{75}{40}
\z{1159}{75}{3}\x{1159}{72}{40}
\end{picture}}
\end{picture}

\newpage\begin{thebibliography}{99}

\bibitem{ddl} D.~B. DeLaney \etal, preprint UTHEP-92-0101, 1992.
\bibitem{sbm} S.~Jadach and B.~F.~L. Ward, \pr{38} (1988) 2897; \ibid\ {\bf
D39}
(1989) 1471; \ibid\ {\bf D40} (1989) 3582; \cpc{56} (1990) 351; CERN-TH
6230-91; \pl{274} (1992) 470, and references therein.
\bibitem{yfs} D.~R. Yennie, S.~C. Frautschi, and H.~Suura, \anp{13} (1961) 379;
K.~T. Mahanthappa, \prz{126} (1962) 329.
\bibitem{prt} M.~Gl\"uck, E.~Reya, and A.~Vogt, Z. Phys. {\bf C53} (1992) 127;
D.~W.~Duke and J.~F.~Owens, \pr{30} (1984) 49.
\bibitem{mueller} V.~N.~Gribov and L.~N.~Lipatov, Sov. J. Nucl. Phys.
{\bf 15} (1972) 438; {\it ibid.} {\bf 15} (1972) 675;
Yu.~L. Dokshitser \etal, Rev. Mod. Phys. {\bf 58}
(1980) 269;Yu.~L.~Dokshitser \etal,{\it Basics of Perturbative QCD},
(Frontieres,Gif-sur-Yvette,1991);R.D.Field,{\it Applications of
Perturbative QCD},(Addison-Wesley,Redwood City,CA,1989), and references
therein.
\bibitem{motsai} L.~W.~Mo and Y.-S.~Tsai,Rev. Mod. Phys. {\bf 41} (1969) 205.
\bibitem{tap} D.~B. DeLaney \etal, to appear.
\end{thebibliography}
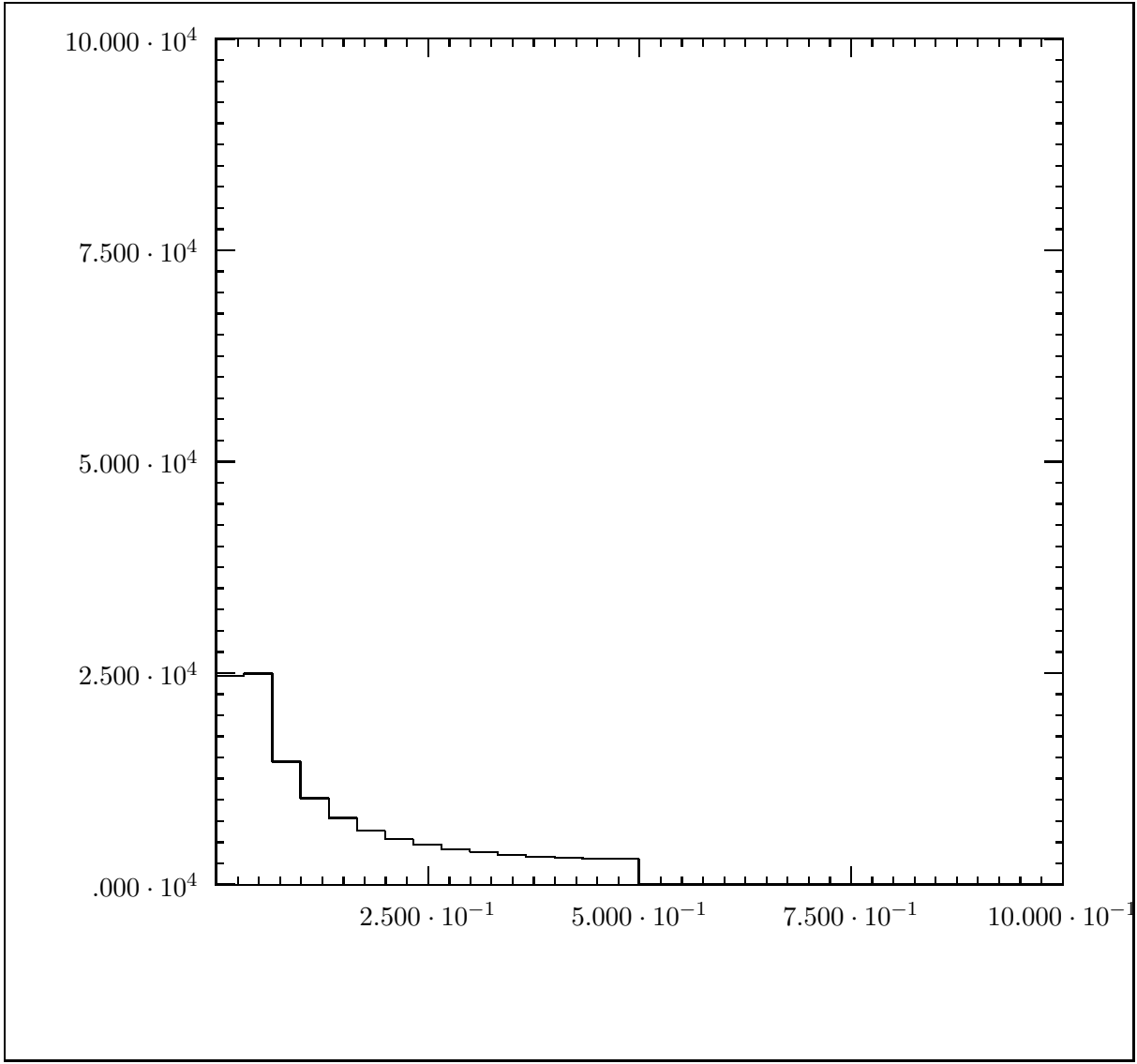
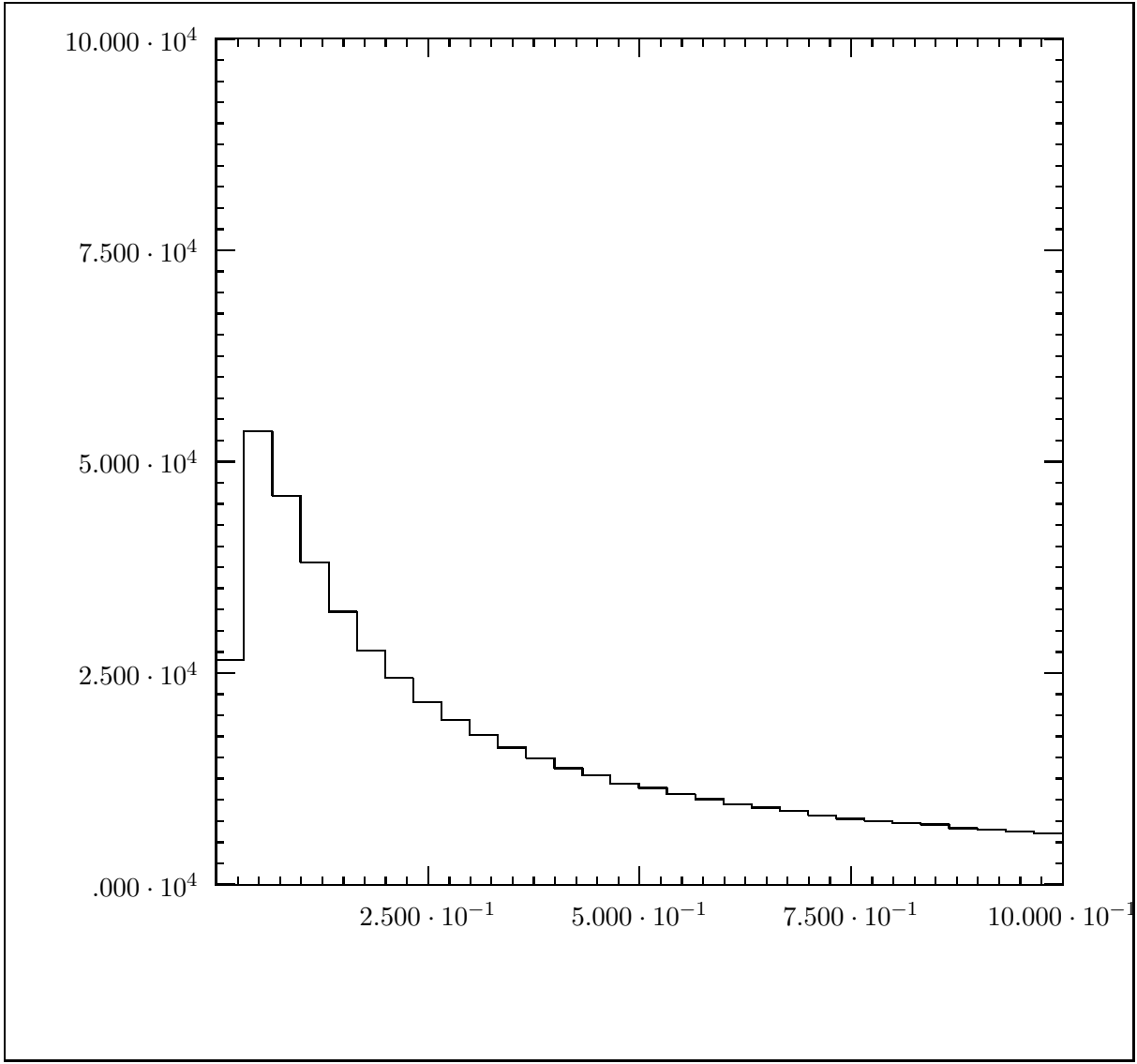
\end{document}